# A Simple Yet High-Performing On-disk Learned Index: Can We Have Our Cake and Eat it Too?

Hai Lan*, Zhifeng Bao*, J. Shane Culpepper*, Renata Borovica-Gajic★, Yu Dong†
*RMIT University, ★The University of Melbourne, †PingCAP

## ABSTRACT

While in-memory learned indexes have shown promising performance as compared to B+-tree, most widely used databases in real applications still rely on disk-based operations. Based on our experiments, we observe that directly applying the existing learned indexes on disk suffers from several drawbacks and cannot outperform a standard B+-tree in most cases. Therefore, in this work we make the first attempt to show how the idea of learned index can benefit the on-disk index by proposing AULID, a fully on-disk updatable learned index that can achieve state-of-the-art performance across multiple workload types. The AULID approach combines the benefits from both traditional indexing techniques and the learned indexes to reduce the I/O cost, the main overhead under disk setting. Specifically, three aspects are taken into consideration in reducing I/O costs: (1) reduce the overhead in updating the index structure; (2) induce shorter paths from root to leaf node; (3) achieve better locality to minimize the number of block reads required to complete a scan. Five principles are proposed to guide the design of AULID which shows remarkable performance gains and meanwhile is easy to implement. Our evaluation shows that AULID has comparable storage costs to a B+-tree and is much smaller than other learned indexes, and AULID is up to 2.11x, 8.63x, 1.72x, 5.51x, and 8.02x more efficient than FITing-tree, PGM, B+-tree, ALEX, and LIPP.

## 1 INTRODUCTION

Nowadays, most widely used database systems still rely on on-disk indexing techniques for (at least) two reasons. First, the total index size may be larger than the main memory available – a consequence of growing data sizes in real applications [1]. Also, multiple indexes (not just one index) might be built to optimize workload-specific performance [40]; they are usually operationalized as a "secondary index", where the leaf nodes should be included when calculating the total storage requirements. Second, main memory is also a precious resource for efficient query processing to store intermediate results, e.g., a hash table in a hash join [9]. If most of the available memory is used to hold the indexes, query performance can be significantly degraded. On the other hand, most existing learned indexes are designed for main memory setting and try to reduce the search/insert overhead via different approaches: (1) use model-based search instead of binary search [4, 35] ($O(1)$ vs. $O(\log n)$); (2) have a smaller search range when employing a binary search [7, 8]; (3) reduce the tree height by modeling the data distribution [7, 35]; (4) support queries using cache-aware techniques [39]; (5) use a gapped array instead of a packed array [4, 35].

When adapting the idea of learned indexes to a fully on-disk setting, most of these techniques are no longer useful since I/O costs are the dominant bottleneck. For example, when issuing a lookup query in a four-layer B+-tree, we find 93.6% of the total execution time is spent on I/O operations. Hence, reducing the number of block reads (and writes) is a critical performance factor.

An immediate question to ask then is how existing learned indexes perform on disk? To answer that, we implemented four state-of-the-art updatable learned indexes [4, 7, 8, 35] on disk and compared them against a standard B+-tree across six workload types commonly encountered by a database. Figures 1(a)-(b) present the normalized throughput on *COVID* and *FB*, which are a representative of easy and hard dataset respectively, as per the dataset profiling in a recent experimental study on learned indexes [34]. We observe that although these learned indexes exhibit different strengths and weaknesses depending on the workload type and dataset distribution, *none of them outperforms or achieves competitive performance to the B+-tree across all workload types on any dataset*. This should come as no surprise to database designers, given that most research on learned indexes has focused on in-memory performance. The benefit of learned indexes in main memory and the shortage of current learned index on disk motivate us to develop a high-performing on-disk learned index. In the rest of this section, we will outline the challenges when building on-disk learned indexes, our design principles to mitigate them, followed by our solutions that align with these principles. Readers can refer to Figure 2 for an overview from challenges to our proposed design principles and solutions.

### 1.1 Challenges in Building a Fully On-disk Learned Index

When (re)implementing existing in-memory learned indexes on disk, several critical challenges discussed next frequently arise. For the reason of providing an intuitive illustration of these challenges, some experimental comparison and analysis are highlighted in between.

*Challenge 1: a learned index cannot guarantee to reduce I/O costs when searching data on disk.* Figure 1(c) shows the average number of inner nodes, inner blocks, and total blocks per query for Lookup-Only and Scan-Only workloads on the *FB* dataset[1]. For LIPP, the total number of fetched nodes is reported, and the number of nodes in the scan is highlighted in the bracket of the fifth row. When combining the results for a Lookup-Only workload and a Scan-Only workload in Figure 1(b), we observe that the performance rank is directly correlated with the number of fetched blocks. In contrast to in-memory indexes, reducing the search overhead for each step does not help on-disk indexes. Instead, reading or writing a block from/to disk is the main overhead. In a Lookup-Only workload, among all the learned indexes, only LIPP fetches fewer blocks than a B+-tree and is more efficient than the B+-tree.

---

[1]For a Scan-Only workload, we set the start key to the same key that was used in the Lookup-Only workload, and then we scan forward 99 keys. This ensures that ALEX,

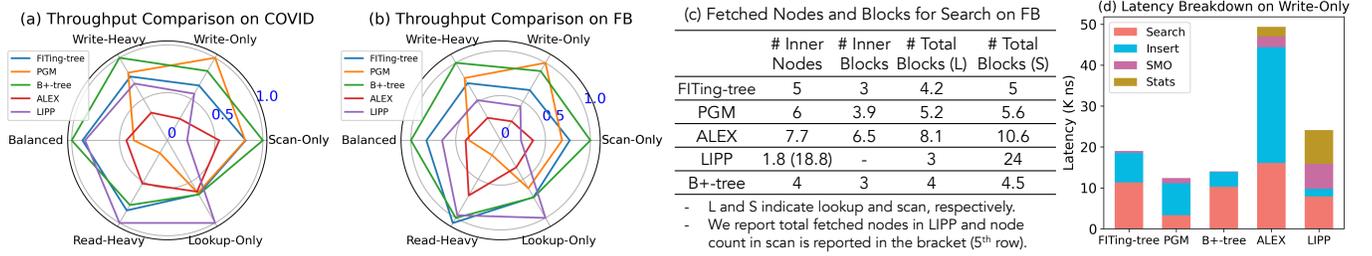

Figure 1: Throughput Comparison and Analysis. Each index's throughput in (a)-(b) is normalized by the largest under the same workload (higher is better), (c) is an analysis on the fetched blocks per query, and (d) is a latency breakdown per query.

We also observe that, in contrast to the B+-tree, existing learned indexes have larger scan overheads, which, in practical terms, means fetching the *next* item becomes more expensive. For example, to support a scan, LIPP traverses many nodes, which incurs a higher I/O cost and leads to poorer performance.

*Challenge 2: large insertion overheads.* Current indexing techniques support an insertion using four steps: (1) find a slot to hold a new key-payload pair (**Search**); (2) do the insertion (**Insert**); (3) induce an index structure modifications operation (**SMO**) if necessary; (4) update various statistics (**Stats**), such as the total number of lookups and insertions, which determine when to induce SMOs. An SMO may create new nodes or re-construct an entire sub-tree during the insertion, which is necessary for the index to allocate empty slots, and for a learned index to benefit from future model-based operations. Figure 1(d) shows the average latency breakdown per query for a Write-Only workload on the *FB* dataset. We observe that learned indexes have several shortcomings. ALEX and LIPP have a large overhead when updating statistics and performing an SMO. The FITing-tree, PGM, and ALEX incur a large overhead for the insertion. LIPP frequently induces an SMO to resolve conflicts between two keys, while ALEX re-writes large nodes (the leaf node) in every SMO. Both lead to large SMO overheads. The shift operations, which are used to obtain an empty slot to store a new key-payload pair in a FITing-tree, PGM, and ALEX, may span multiple blocks on disk – leading to more writes on disk.

### 1.2 AULID – Simple Is Better

In this paper, we show that the idea of learned index can benefit the on-disk index design by proposing AULID, <u>a</u>n <u>u</u>pdatable <u>l</u>earned <u>i</u>ndex on <u>d</u>isk. To address the above challenges, we propose five design principles for such an on-disk learned index in Section 3.1, directed toward supporting lower tree heights and lower SMO overheads. Then, we design AULID to meet these principles by

---

PGM, a FITing-tree, and a B+-tree fetch the same number of inner nodes and blocks needed for a lookup and a scan.

leveraging the idea of B+-tree and learned indexes in three ways (an overview from the challenges to our design principles and solutions can be found in Figure 2):

❶ **Leaf Node Layout.** To reduce the insertion overheads, instead of using a learned model to search items for all layers, we use model-based search only for inner nodes. This helps reduce the burden on leaf nodes in maintaining the benefits of model-based search. Specifically, we use a B+-tree styled layout for leaf nodes, which has a low overhead when updating the index. Since the majority of SMOs are on leaf nodes, a lightweight SMO mechanism for leaf nodes, as achieved with AULID, reduces the insertion overheads significantly. A B+-tree styled leaf node design also benefits scan operations in fetching the *next* item.

❷ **Inner Node Layout.** After building a B+-tree styled leaf node, the path from the root node to a leaf node should be shorter than that in a B+-tree. Otherwise, the learned index cannot outperform a B+-tree for on-disk operations. The results in Figure 1(a)-(b) inspired us to adopt the Fastest Minimum Conflict Degree (FMCD) algorithm in LIPP [35] to reduce the tree depth. Although it is not the best indexing method for most workload types, it has the smallest number of fetched blocks for a lookup, making it suitable for on-disk indexes when attempting to reduce I/O costs. Moreover, a lookup is often the first step in other operations. For example, in a scan operation to find the position of a start key and in an insert operation to find a position to hold a new key-payload pair. Thus, a better performance of the lookup operation should boost the performance of other operations too. However, for certain workload types and datasets, e.g., *OSM* (a hard dataset) [34], directly applying the aforementioned inner node layout still reveals several shortcomings – a larger storage size and a lower throughput. To overcome these shortcomings, we introduce two new inner node types and design processing algorithms upon the new layout. For example, with our design, AULID is 1.18x more efficient on the Lookup-Only workload while only taking 0.84x storage on *OSM* (as compared to a B+-tree).

❸ **Structural Modification Operations.** With a B+-tree styled leaf node, AULID already manages to achieve a lighter overhead in modifying the index structure due to a lower frequency in updating the inner nodes and a lighter overhead in updating the leaf node. However, the tree height in some region could grow and even become larger than that of a standard B+-tree. In turn, AULID will have a worse performance after lots of insertions. To avoid that from happening, we monitor the tree height of each branch and trigger a re-construct process to bound their tree height if needed.

To summarize, we make the following technical contributions:
- To the best of our knowledge, AULID is the first approach to employ the ideas of learned index to a fully on-disk setting to replace a traditional B+-tree[2].
- We propose five principles to guide the design of an on-disk updatable learned index (Section 3.1), and carefully design the indexing structure (Section 3.2 - 3.3), the query processing algorithms (Section 4.1 - 4.3), and an SMO mechanism, to achieve efficient reads and writes (Section 4.4).
- We implement AULID in C++, conduct comprehensive experiments across a wide range of datasets and workloads and compare it against a B+-tree and on-disk implementations of existing in-memory learned indexes. Our evaluation shows that AULID has competitive storage costs to a B+-tree and is much smaller than most other learned indexes. Performance-wise, AULID achieves up to 2.11x, 8.63x, 1.72x, 5.51x, and 8.02x larger throughput than FITing-tree, PGM, B+-tree, ALEX, and LIPP, respectively. We also conduct an in-depth evaluation on the benefits of the AULID design (Section 5).

## 2 RELATED WORK

Given that our work focuses on developing a fully on-disk learned index under the single-threaded setting, we start the literature review on learned indexes outside main memory, followed by an overview of in-memory ones and a discussion on the difference in concurrency support between in-memory case and on-disk case.

**Learned Indexes Outside Main Memory.** The authors of [1, 3] studied how to use learned indexes on disk in a log-structured merge tree (LSM) [23]. A learned model is constructed for each Sorted Strings Table (SSTable), which is *immutable* after being created. Modification operations (insert, update, delete) are supported in an LSM framework. Models are *rebuilt* and dropped during periodic compaction processes. The LSM framework supports efficient writes at the cost of reads. In contrast, our work focuses on building an on-disk learned index with the hope of replacing the B+-tree.

Two most recent studies [19, 38] focus on the *larger than main memory setting*, where they pin part of the index in main memory and introduce different caching strategies. TreeLine [38] uses the partitioning algorithm proposed in PGM [7] to generate the leaf nodes and adopts a B+-tree to index them. The B+-tree in TreeLine is pinned in main memory, and a record-level caching strategy is used to cache frequently accessed items in main memory. FILM [19] builds a PGM index, stores it in main memory and uses one bit for each item to indicate the location of that item, i.e., in main memory or on disk. Moreover, FILM introduces a global chain and a local chain, to organize the segments at the last level and the items in the segment based on their access time, respectively. In this way, FILM can quickly locate cold items. However, FILM is designed for *append-only insertions*. Differently, our work focuses on storing the whole index on disk rather than only leaf nodes.

Lu et al. [18] propose APEX, a learned index for persistent memory (PM) [33]. APEX is a variant of ALEX, with several tailored designs: (1) different node size settings used in APEX – a larger inner node and a smaller data node to reduce SMO overhead; (2) a new probe-and-stash mechanism to resolve collisions without introducing unnecessary nodes' access; (3) concurrency control and recovery mechanisms are introduced to support simultaneous inserts and instant recovery. Different from APEX, on-disk operations are our focus where I/O costs are the main overhead.

**Learned Indexes in Main Memory.** Kraska et al. [14] was the first group to propose the idea of learned index, where a hierarchy of models, called RMI, was built to replace a B+ Tree for sorted 1-d data. The new approach can achieve 3x performance boost and 10x smaller index size. Given that RMI only supports lookup queries, subsequent studies [4, 7, 8, 35] address this limitation using tailored index structures and new mechanisms for index structure modification. A FITing-tree [8] replaces the last layer of a B+tree with model-based search, and supports insertions by introducing buffers for each segment. PGM [7] uses a similar idea to the FITing-tree, but it leverages model-based search for every layer based on an optimal partitioning algorithm [24]. PGM handles arbitrary insertions in an LSM tree [23] manner. Although the FITing-tree and PGM leverage model-based search, additional binary search operations are needed. ALEX [4] inserts a key-payload pair using model prediction, and hence manages to have accurate predictions in the inner nodes without binary search. At each data node (leaf node), ALEX uses a gapped array, where the key-payload pairs and empty slots are interleaved. The gapped array can reduce the frequency of shifts for new insertions. Without an error bound for the search range, ALEX uses an exponential search to find the target position. Model prediction in LIPP [35] tends to be accurate in every layer. LIPP has been shown to have better performance in practice than other learned indexes in most settings [34]. However, LIPP requires much more memory and is not efficient for a range query. Wu et al. [36] use Normalizing Flows [28] to transform a dataset so that it can be easily modeled with linear models, and extend the idea of LIPP by introducing a bucket node type. CARMI [39] is a cache-aware learned index which uses tailored partitioning algorithms and a prefetching mechanism for the in-memory setting. To verify the efficiency of learned indexes, several studies have conducted comprehensive experimental studies [21, 34], detailed theoretical analysis [6], and performance analysis [20].

There are also several studies that propose read-only learned indexes [11, 13, 30] or use the model to boost B+-tree's performance [10, 17] for main memory operations. Several other learned indexes have also been proposed as secondary indexes [12, 37], tailored for multi-dimensional data [2, 5, 22], spatial data [16, 25–27], or string data [29, 32].

**Learned Index With Concurrency Support.** Among these in-memory learned indexes, the XIndex [31], FINEdex [15], ALEX+[34], and LIPP+[34] support concurrent operations. All of them use optimistic locking, which associates a versioning lock at each node. With the node size being larger than a block at the leaf node, TreeLine [38] proposes some locking strategies based on the node-level lock and block-level lock to support concurrent operations on the LSM data structure. Aligned with the latest study on the larger than main memory case [19], we focus on the single-threaded setting in this work. In a fully on-disk learned index with different sizes of node and block, the node-level lock and the block-level lock need

---

[2]Two recent studies [1, 3] on disk are built upon LSM tree [23] and suffer from poor read performance. Details are presented in Section 2.

, , Hai Lan*, Zhifeng Bao*, J. Shane Culpepper*, Renata Borovica-Gajic★, Yu Dong†

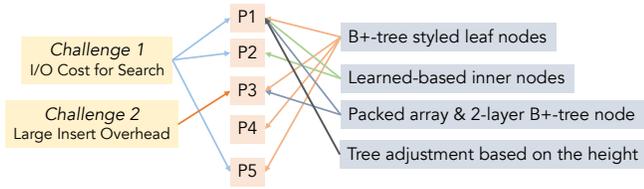

Figure 2: An Overview of AULID: from Challenges to Design Principles and Solutions

to be introduced at the same time. Without a tailored mechanism, one cannot achieve a good scalability in a multi-core setting due to the block-level lock. Although TreeLine [38] has introduced these two locks, its inner nodes are pinned in main memory, which is easier than the above case. We believe our work is an important first step and our findings will help the community in the future, when designing a fully on-disk concurrent learned index.

## 3 AN OVERVIEW OF AULID

In this section, we first introduce the principles and highlights of the AULID design – addressing the challenges discussed in Section 1.2. Then, we present the AULID layout. Figure 2 illustrates two identified challenges and the associated design principles used to resolve them, as well as how these principles are reflected in our proposed solutions.

### 3.1 Design Principles

Based on the key properties arising from disk and learned indexes, we propose a number of principles to guide the design of AULID:

- **P1. Reducing the Tree Height of the Index.** Accessing each level in an index requires at least one disk access when an index is stored on disk. Reducing the tree height can reduce the number of disk access.
- **P2. Model-based Operations (Search and Insert).** An index with a reduced height usually has larger nodes in certain levels of the index. Model-based operations help AULID quickly find search keys in a specific part of the node, without the need to access the entire node on disk.
- **P3. Lightweight Structure Modification Operations.** Structure modification operations (SMOs) for the existing learned indexes incur a substantial amount of writes on disk. AULID should reduce the overhead of such SMO calls.
- **P4. Support Duplicate Index Keys.** Duplicate (i.e. non-unique) index keys are common in real systems. Typically, they can be supported using a linked list in a main memory setting [35], but not on disk, since it leads to additional disk reads.
- **P5. Better Scan Performance.** Existing learned indexes have their own limitations when supporting scans on disk (see Figure 1(a)-(b)).AULID must provide a lightweight method to fetch the *next* item efficiently.

### 3.2 Design Highlights

AULID uses a combination of existing and novel techniques to meet the above principles and achieve high performance on disk. AULID consists of inner nodes and leaf nodes, both of which are stored on disk. Leaf nodes, where most SMOs occur, are organized in a B+-tree manner. A low update cost at leaf nodes reduces the SMO overhead (**P3**). Moreover, AULID only uses the idea of learned index for inner nodes to index the maximum key of each leaf node, which leads to less frequent SMOs in updating the inner nodes (**P3**) and a low tree height in inner part (**P1**). Each leaf node is a packed array – it stores pointers to its siblings and its size is equal to the *block size*. Using the packed array and links to siblings, AULID can support efficient scan operations (**P5**). We optimize our inner nodes based on properties of the disk drive. Fast lookup time with the learned model means that AULID can efficiently locate target leaf nodes (**P1, P2**). To achieve robust performance on different datasets (i.e., different distributions), we also introduce two new node types for the inner nodes, a packed array and a two-layer B+-tree, with the purpose of reducing the number of SMOs for non-leaf nodes (**P3**). By proposing a tailored mechanism to handle duplicate keys inserted in inner nodes, AULID manages to store duplicate values with reduced on-disk costs (**P4**). To maintain the performance gains achieved from the learned model, AULID adjusts the index structure based on the tree height and bounds the tree height during insertions (**P1**).

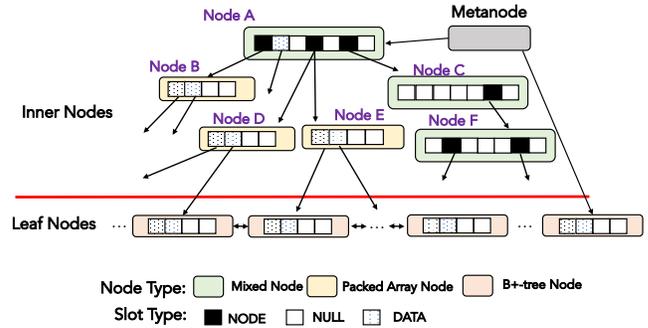

Figure 3: AULID Index Structure

### 3.3 Node Structure

The index structure of AULID is presented in Figure 3. Similar to existing indexes, AULID is composed of two components: the inner nodes which store the route information to leaf nodes, and the leaf nodes which store the key-payload pairs.

*3.3.1 Metanode.* Metanode in AULID stores (1) the physical address of the root node, (2) the linear model of the root node, and (3) the physical address of the last leaf node, as well as the minimum and maximum keys of that node. We store the metanode in main memory, which requires only 80 bytes, a negligible main memory overhead.

*3.3.2 Inner Nodes.* AULID has two node types in the inner part, a mixed node type and a packed array node type. And there are three types of slot in the inner part: *NODE*, *NULL*, and *DATA*. The *NODE* slot stores the pointer to its child. The *NULL* slot is the empty slot and can be converted to *NODE* or *DATA*. The *DATA* slot stores the key-payload pair. Each mixed node has a model to predict the slots for a key search and can include three different slot types above. AULID stores the model in the parent node, combined with the physical address. If we store the model at the starting address of a mixed node, the large fanout for mixed nodes increases the chance that the predicted position and the model are located in different blocks. Thus, two blocks must be fetched from disk for each level in the tree. In contrast, when storing the model in the parent node, AULID only fetches one block per level.

The *NODE* slot in AULID can be further divided into three types: (1) a pointer to the packed array of fixed size. That is, the first slot in Node A in Figure 3; (2) a small B+-tree node that contains at most four child nodes – the fourth slot in Node A; (3) a pointer to another mixed node – the sixth slot in Node A. For the first case, we introduce four different packed array types, each with a fixed size. The $i^{th}$ packed array type can store $2^{i+2}$ items of *DATA* types, where $i$ ranges from 1 to 4. A *DATA* slot in the inner nodes stores the physical address of a leaf node and the largest key it contains, i.e., the key-block pair. The second case is proposed to improve the performance for scenarios where the number of keys to be inserted into the same slot is greater than 64, but smaller than 1020 (which will be explained later in Section 3.3.2). This indicates the *conflict degree* for the region. If we create a new mixed node to hold these keys, there can be key conflicts in the new node. This leads to a larger tree height (more than two layers). Conversely, a two-layer B+-tree can be used to hold the nodes and help AULID to bound the tree height for the region. Therefore, AULID is able to bound the number of fetched blocks. Also, the design of the packed array and the two-layer B+tree can support more newly inserted key-block pairs (the routing knowledge to the leaf nodes) to be stored with low overheads. Using the packed array and a two-layer B+-tree, AULID achieves a better empty slot ratio, and this translates to smaller storage costs.

The B+-tree node contains only four child nodes for the following reasons: (1) The conflict degrees in most of the test datasets we have used (except for one)[3] is less than 1000, which can be stored easily using a two-layer B+-tree. (2) A larger fanout requires more metadata (pivot keys and physical addresses) in a slot, and it increases the total storage cost significantly. A key-block pair occupies 16 bytes on disk in our implementation. Thus, a block with 4 KB can store 256 pairs. The first item records the item count for a two-layer B+-tree's leaf node. Four children can store at most 1020 items.

*3.3.3 Leaf Nodes.* The leaf nodes have the same structure as a standard B+-tree. The *DATA* slot in the leaf node stores the key-payload pairs to be indexed. This layout design is based on the observation that most SMOs happen on leaf nodes as new key-payload pairs are added. Learned indexes need to read all of the items in a large leaf node and re-write them to disk to maintain the benefits of their unique structure, which incurs large I/O costs on disk. A lightweight SMO overhead for a leaf node design can help significantly reduce the number of SMO operations required (see the experiments in Section 5.2.2).

This simple yet elegant design in the leaf nodes has many other benefits. First, the link between siblings when using a packed storage layout requires no additional utility structures to perform efficiently when scan operations must locate the start of a query range. Second, the storage costs of the inner nodes can be significantly reduced by only storing the largest keys. In our experiments,

---
[3]We have also tested all of the datasets proposed in a recent benchmark paper [34].

AULID has a similar storage size and bulkload time as a B+-tree on disk, which is better than other learned indexes. Third, reducing the items inserted into the inner nodes also decreases the SMO frequency and the number of items that must be processed. Last, AULID is able to efficiently support duplicate index keys when using a B+-tree styled leaf node.

## 4 AULID OPERATIONS

First, we present how AULID supports each type of operation, and then we discuss how structural modifications are supported.

### 4.1 Bulkload

AULID supports bulkload using two steps. In the first step, it creates leaf nodes to store the key-payload pairs using B+-tree styled leaf nodes. When building leaf nodes – with the exception on the last leaf node – AULID records the maximum key, and the physical address for each leaf node, i.e., the key-block pairs to be indexed in the inner nodes. For the final leaf node, AULID stores the minimum and maximum keys, as well as its physical address in a meta-node.

The second step builds the inner nodes for AULID over the key-block pairs. We first use the Fastest Minimum Conflict Degree (FMCD) algorithm in LIPP to generate a linear model for a node. Given the collection of keys to be indexed and the number of slots that can be used, FMCD aims to generate a linear model under which the maximum number of keys inserted into the same slot is minimized, i.e., the smallest "conflict degree". Then, we insert the key using the resulting model. If only one key is inserted into a slot, this slot is labeled as *DATA* and used to store the key-block pair. Different from LIPP, AULID does not aggressively create a new node if more than one item is mapped to the same slot. Instead, we divide them into three cases depending on the size of the items that are mapped into the same slot: (1) If the size of the items mapped to one slot is smaller than 64, a packed array is created. (2) If the size is greater than 64 and less than 1020 (see explanation at the end of Section 3.3.2), a two-layer B+-tree is created with at most four child nodes. (3) Otherwise, a new mixed node is created to hold the keys.

### 4.2 Lookup & Scan

*4.2.1 Lookup.* Given a search key, we first check whether it belongs to the last leaf node by comparing it with the minimum key and the maximum key that are stored in the meta-node. The overhead of this operation is negligible as the meta-node resides in main memory. If the key belongs to the last leaf node, the leaf node is read from disk, and then a binary search is initiated. Otherwise, the inner nodes are searched to find the leaf node address where the search key should reside.

When traversing from root node to leaf node, five different cases of model prediction can occur (assume a mixed node is the root node):

- *DATA* Slot: A leaf node is fetched based on the physical address contained in it. If the key in the *DATA* slot is less than the search key, then we fetch the successor.
- *NODE* Slot for a Packed Array: The packed array content is retrieved from disk, and a *DATA* slot is located to hold the search key. It is then processed in the same way as the *DATA* Slot case.

- *NODE* Slot for a B+-tree: Just as in a standard B+-tree, a child node is found which holds the search key (if it exists), and then it is fetched and processed in the same way as the *NODE* Slot for a Packed Array.
- *NODE* Slot for another mixed Node: The model from the node is used to predict which node to access next, and the search process is repeated.
- *NULL* Slot: Using the monotonic linear function from AULID and indexing the largest keys for each leaf node, to find the next *DATA* slot we must search forward. For example, given a search key, suppose the predicted position is the $5^{th}$ slot in *Node A* of Figure 3, which is a *NULL* slot. AULID will scan forward to find the *next DATA* slot using *Node C* and *Node F*.

*4.2.2 Scan.* Given a query range $[u, v]$, we first call a lookup operation to locate the leaf node where $u$ should reside, and the position of $u$ in the node. Then, we scan forward until reaching the last key $v$. Using the links to sibling leaf nodes and the packed array, the *next* item can be quickly accessed without any additional utility structures, such as bitmaps in ALEX to differentiate empty slot, or traversing many nodes in LIPP.

*4.2.3 Optimization for Reading.* When storing only the largest key of each leaf node in the inner nodes, AULID could issue additional I/O requests for two reasons:

- *Issue 1*: When traversing the inner nodes, AULID may need to find the predecessor (the left sibling of a target node) and extra I/O is needed to fetch that target.
- *Issue 2*: If the predicted location is a *NULL* slot, a scan operation is triggered to locate the *next DATA* slot. For example, if the predicted position is the $5^{th}$ slot in *Node A* of Figure 3, AULID needs to access *Node C* and *Node F*. Due to the large fanout of the inner nodes, *Nodes A*, *C*, and *F* are stored in different blocks, which could incur additional I/O costs. Note that with a monotonic linear function in AULID, these two cases cannot happen in a single lookup at the same time.

To address the first issue, if a *DATA* slot is found, instead of fetching the leaf node directly, we first check whether the key it contains is smaller than the search key. If so, we scan forward to find the next *DATA* slot. In contrast to a *NULL* slot, the scan operation is initiated only on the currently fetched block. The *DATA* slot found before is used if no new *DATA* slots are found in the same block. Otherwise, scanning forwards reads at least one additional block.

To address the second issue, AULID fulfills the preceding *NULL* slots for one *DATA* slot until reaching the previous *DATA* node during the bulkloading process. This operation has negligible overhead for the bulkloading process while it only works with Read-Only workloads. For any workload involving a write operation, inserting a new key-block pair into the inner nodes will incur an update for empty slots and hence leads to increased latency during an insertion. The effectiveness of these two optimizations have also been verified in our evaluation at Section 5.4.

### 4.3 Insert & Duplicate Index Keys Support

*4.3.1 Insert.* The full insertion process is presented in Algorithm 1, and the key process is depicted in Figure 4. Given a new key-payload pair $(k, p)$, a lookup operation is first called to locate the leaf node

**Algorithm 1:** Insert($\mathcal{I}$, $k$, $p$)

**Input:** $\mathcal{I}$: the AULID index, $k$: the key, $p$: the payload, $T$: maximum slot count in the leaf node

1 accessed_nodes = [];
2 leaf_node = GetLeafNode($\mathcal{I}$, $k$);
3 **if** leaf_node.size $\geq T$ **then**
4     block_id, $k_{max}$ = SplitNode(leaf_node); ▸ max key and block id of left child;
5     $e$ = FindEntry($\mathcal{I}$, $k_{max}$, accessed_nodes); ▸ first non mixed Node entry;
6     **switch** $e.type$ **do**
7         **case** *NULL* **do**
8             insert ($k_{max}$, block_id) into $e$;
9         **case** *DATA* **do**
10            create a packed array *PA*;
11            insert ($k_{max}$, block_id) and (DATA.k, DATA.v) into *PA*;
12            update $e$ as *NODE* to *PA*;
13         **case** *B+-tree* **do**
14            **if** $e$ is full **then**
15                $K$ = ($k_{max}$, block_id) $\bigcup$ items in $e$;
16                $P$ = BuildMixedNode($K$); ▸ Build a mixed node;
17                update $e$ as *NODE* to $P$;
18            **else**
19                insert ($k_{max}$, block_id) into $e$ ;
20         **case** *Packed Array* **do**
21            **if** $e$ is full **then**
22                create a packed array or a B+-tree node $P$;
23                insert ($k_{max}$, block_id) and items in $e$ into $P$;
24                update $e$ as *NODE* to $P$;
25            **else**
26                insert ($k_{max}$, block_id) into $e$ ;
27     StatsUpdate(accessed_nodes); ▸ Update statistics for SMOs;
28     Adjust($\mathcal{T}$, accessed_nodes); ▸ See Algorithm 2;
29 **else**
30     InsertLeaf($k$, $p$, leaf_node);

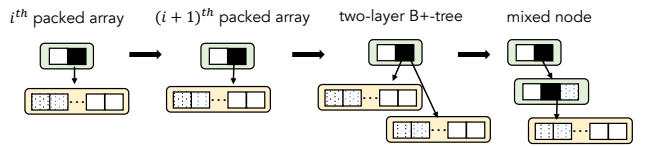

**Figure 4: AULID 's Insertion Process**

that should contain $k$ and the position of $k$ in the node (Line 2). If the item count in this leaf node is less than a predefined threshold $T$, $(k, p)$ is added into this node (Line 30). Note that, if this is the last leaf node, and $(k, p)$ was the first or last item in the node, the minimum key or maximum key is updated in the meta-node. If the item count exceeds the threshold, a splitting process is triggered (Line 4). Unlike a B+-tree, which keeps the smaller half of the items in the original node, AULID keeps the larger half of the items in the original node. Otherwise, the address of the last key is updated in the original inner node, which requires extra writes. After a new leaf node is created, the links to the sibling nodes are also updated.

**Algorithm 2:** Adjust($\mathcal{I}$, $N$)

**Input:** $\mathcal{I}$: the AULID index, $N$: accessed mixed nodes

1 **foreach** $n \in N$ **do**
2   **if** $n.size \geq \beta \cdot n.init\_size \wedge n.l3\_item \geq \alpha \cdot n.size$ **then**
3     $K$ = items (inner part only) in the subtree rooted as $n$;
4     $P = BuildMixedNode(K)$;
5     update the pointer to $n$ in its parent to point to $P$;
6     *StatsUpdate*(ancestors of $P$);  ▸ This step can be merged with Line 27 in Algorithm 1 to reduce the write overhead;

After the leaf node is generated, it is indexed (the largest key and physical address) in the inner nodes (Lines 5-28). A lookup process is initiated to find the first non-mixed node slot to hold the new key. Just as in the search process, there are five different cases: 1) If we encounter a *NULL* slot, we insert the new key-block pair into it and the insertion is completed (Line 8). 2) If we encounter a *NODE* slot pointing to a mixed node, the model is fetched and the search process is repeated. 3) For a packed array, if full, a larger packed array type will be allocated. If the maximum supported packed array is already being used, it is converted into a two-layer B+-tree (Line 20-24). Otherwise, we insert the new key-payload pair into the empty slot and complete the process (Line 26). 4) If the B+-tree is not full, the process proceeds as in a standard B+-tree (Line 19). 5) Otherwise, it is converted into a mixed node (Lines 15-17).

After completing the insertion, the statistics of the mixed nodes (*accessed_nodes*) are updated in the access path to guide later SMOs (Line 27). AULID records the number of items in a third layer or a deeper layer. Finally, we check if we need to initiate an SMO operation by calling the *Adjust* function (Line 28).

*4.3.2 Handling Duplicate Keys.* Duplicate keys are common in real databases. A linked list used in main memory is however not appropriate when on-disk, as it leads to additional I/O costs when fetching items from the list.

Using a B+-tree styled leaf node, AULID can efficiently store and search for duplicate keys in the leaf nodes. This process is the same as a standard B+-tree. If a duplicate key must be inserted into the inner nodes of AULID, one potential way is to directly insert the duplicate key into the inner nodes. AULID can handle key conflicts using the packed array/two-layer B+-tree proposed earlier. However, in this case, the maximum number of duplicate keys that can be supported is 163, 840 for a block size of 4 KB[4]. Using the link between two sibling leaf nodes, another way is that we can only store the first leaf node's address for the duplicate key. With the larger half of items stored in the original block after the leaf node is split, the address (block number) stored in the original slot in the inner nodes needs to be updated. The write overhead is the same as the last case while we can support an arbitrary number of duplicate keys.

### 4.4 Structural Modification of Inner Nodes

As shown in Figure 4, in AULID, inserting new key-block pairs into the packed array and the two-layer B+-tree will not increase the tree height, i.e., not incur an additional I/O request. However, when a two-layer B+-tree node is converted into a new mixed node, certain regions may have a larger tree height. In our test datasets with up to 800M key-payload pairs, a B+-tree has at most three layers in the inner nodes. Therefore, the index structure must be carefully modified to bound the height of the branches in AULID's inner nodes to at most three layers. Otherwise, a B+-tree on disk will be the best. Packed arrays and two-layer B+-tree nodes do not have an impact on the tree height. Here, we focus on when to re-construct mixed nodes in AULID and how to perform a re-construction.

*4.4.1 When Should the Rebuilding Occur?* To bound the tree height of the inner nodes and avoid aggressive node updates, we introduce two new constraints to determine when a mixed node (Line 2 in Algorithm 2) should be reconstructed.

**Criterion 1**: the percentage of the items in a subtree rooted at node $n$ in the third layer or a deeper layer (*l3_item*) is larger than $\alpha$. This guarantees that no more than $\alpha$ leaf nodes can have a longer path than a B+-tree on disk, with a high probability.

**Criterion 2**: The number of current items rooted at node $n$ is larger than $\beta$ times the initial size.

In corner cases where a region has a high degree of conflict, a mixed node can have more than $\alpha$ items, even when it is initially being created. To avoid reconstructing this node frequently, we adjust it after observing a sufficient number of new items.

A smaller value for $\alpha$ and $\beta$ leads to a more frequent node reconstruction. By default, we set $\alpha = 0.05$ and $\beta = 1.2$ to balance the tree height, i.e., lookup performance and SMO overheads. We have verified the impact of different settings for $\alpha$ and $\beta$ in Section 5.4.3.

*4.4.2 How to Reconstruct a Node.* If a mixed node meets both of the above criteria, all key-payload pairs stored in the inner nodes rooted at that node are collected, and then the bulkload process is called again to build a new mixed node.

### 4.5 Other Operations

To support a delete operation, AULID first locates the items to be deleted at the leaf node, and then deletes it in the same manner as a standard B+-tree. If no SMO is required (merging the sibling nodes), the delete operation is finalized. In this case, even if we delete the last key-payload pair in the leaf node, AULID still does not update the inner nodes. If a merge is required, a delete operation in the inner nodes is required. If the key-block pair to be deleted is in a mixed node, this slot is marked as an empty slot. If it is contained in the packed array or a two-layer B+-tree node, it will be removed. AULID will convert the packed array or a two-layer B+-tree node into a *DATA* slot if there is only one key-block pair remaining.

There are two types of update operations, updating the payload and updating the key. In the former, an in-place update is used[5]. For the latter, a delete operation and an insert operation are initiated.

### 5 EXPERIMENTS

We have conducted extensive experiments to answer the following questions:

**Q1:** How good is AULID as compared to other learned indexes and a B+-tree when disk-resident?

---

[4]A B+-tree node with 4 children can store at most 640 leaf node addresses and each leaf node can store 256 items.

[5]Currently, AULID supports key-payload pairs of a fixed length.

, , Hai Lan*, Zhifeng Bao*, J. Shane Culpepper*, Renata Borovica-Gajic★, Yu Dong†

**Q2:** How well does AULID scale to large datasets?
**Q3:** Do the proposed index structure design and structural modification operation help improve the performance?

We start with the experimental setup as described in Section 5.1. Then, we present our answers to **Q1-Q3**. To answer **Q1**, we compare AULID against five competitors across six different workload types and four different datasets. We demonstrate that AULID is superior in terms of throughput and storage cost in Section 5.2. To answer **Q2**, in Section 5.3 we use another four datasets with 800M keys to study the performance of AULID on large datasets of varying hardness. Notably, most existing studies [4, 19, 35, 38] use at most 200M key-payload pairs to study index performance. Finally, to answer **Q3**, in Section 5.4 we compare AULID to its variants, with and without the proposed data structures and structure modification operations, in order to reveal the performance benefits of our proposed design choices.

## 5.1 Experimental Setup

*5.1.1 Baselines.* We implement a standard B+-tree and four state-of-the-art (SOTA) updatable in-memory learned indexes–PGM [7], FITing-tree [8], ALEX [4], and LIPP [35]– all modified to work on disk. To improve the FITing-Tree's performance and reduce the segment count, we replace the greedy partitioning algorithm in the FITing-Tree with a streaming algorithm [24] originally used in PGM. Additionally, to support arbitrary insertions, we support a *Delta Insert Strategy* [8] in the FITing-tree, which allocates a buffer for each segment. LIPP and ALEX use their default settings. For PGM and FITing-tree, we set the error bound as 64, where they achieve good performance in most test cases. PGM supports the insertion operation via the same mechanism as the studies [1, 3], and hence PGM can also reflect their pros and cons.

*5.1.2 Datasets.* The most recent experimental study [34] on memory-resident learned indexes introduced 11 real datasets in their evaluation. Based on its profiling results, these datasets can be roughly divided into four categories (see Figure 2 in [34]): *C1*: Globally easy and locally easy, *C2*: globally normal and locally normal, *C3*: globally normal and locally hard, and *C4*: globally hard and locally normal. We select one dataset from each of these categories for our experiments: *COVID* (*C1*), *PLANET* (*C2*), *GENOME* (*C3*), and *OSM* (*C4*). Each dataset has 200M keys of type *uint64*. The performance of AULID and LIPP correlates to the *conflict degree* (the maximum number of keys being inserted into the same slot) in one dataset due to the usage of the FMCD algorithm. Consequently, datasets with a greater *conflict degree* are more challenging for AULID and LIPP. A summary of the conflict degrees of the tested datasets is presented in Table 1.

**Table 1: Conflict Degree of Each Dataset**

| Dataset | COVID | PLANET | GENOME | OSM |
|---|---|---|---|---|
| *Conflict Degree* | 27 | 22 | 585 | 4,106 |

To test the scalability of AULID on large datasets, we use *OSM* from [21], which contains 800M *uint64* keys. The generator proposed in [34] is used to generate another three datasets, each of which has different levels of hardness (details presented in Section 5.3). All of the generated datasets contain 800M *uint64* keys.

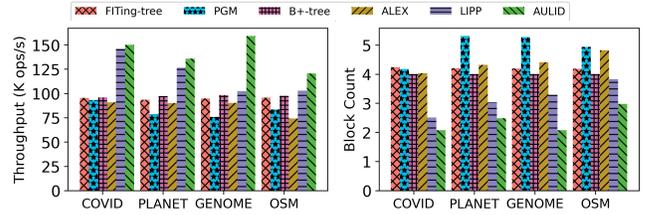

Figure 5: Throughput on Lookup-Only Workload (W1).

It is worth highlighting that, as compared to the latest work on the larger than the main memory setting [19, 38] where at most 30M keys are used, the number of keys in each dataset used in our experiment is much larger.

For all datasets, we generate a *uint64* payload for each key with key plus 1 as their value. The first four datasets require 2.98 GB of storage space on disk, and the last four datasets (used for the scalability testing) require 11.92 GB.

*5.1.3 Workloads.* We compare AULID against all baselines across six different workload types typically encountered in a database.
**W1 - Lookup-Only** workload, where each index is built on 200M key-payload pairs and the workload contains 20,000 randomly sampled search keys.
**W2 - Scan-Only** workload, where the start key of the search range is set to the same key in the lookup-only workload, and the search range is set to 100. The queries are issued on indexes prebuilt on the full dataset.
**W3 - Write-Only** workload, where the initial index is built with 10M key-payload pairs that are randomly selected from a dataset, and then another 10M key-payload pairs are inserted.
**W4 - Read-Heavy** workload includes 90% lookup queries and 10% write operations.
**W5 - Balanced** workload consists of 50% lookup queries and 50% write operations.
**W6 - Write-Heavy** workload includes 90% write operations and 10% lookup queries.

We refer to W4-W6 as mixed workloads, with the only difference between them being the ratio between reads and writes. For mixed workloads, the initial index is built over 10M key-payload pairs randomly sampled from a dataset, and then lookup queries and write operations are issued (10M queries in total). The search keys in all mixed workloads are randomly sampled from the existing keys of an index.

*5.1.4 Metrics & Environment.* The primary metric we measure is throughput. We also report the number of fetched blocks, the storage size of each index, and the tail latency. We conduct the experiments on a SATA HDD using a Red Hat Enterprise Server 7.9 on an Intel Xeon CPU E5-2690 v3 @ 2.60GHz with 256 GB memory and a 1TB HDD. The block size is 4 KB in all experiments.

## 5.2 Efficiency Comparisons on Disk

In this section, we compare AULID against four state-of-the-art learned indexes, and a B+-tree on disk. AULID outperforms all five indexes on every dataset and workload tested.

*5.2.1 Lookup-Only Workload.* Figure 5 shows the throughput and the average number of fetched blocks per query, for each index. Overall, AULID is the most efficient indexing method. Specifically, it achieves up to 1.68x, 2.10x, 1.62x, 1.76x, and 1.55x higher throughput than the FITing-tree, PGM, B+-tree, ALEX, and LIPP, respectively. LIPP is the second most efficient index across the majority of datasets. The performance of each index is directly correlated to the number of fetched blocks where more fetched blocks lead to a lower throughput. The FITing-tree, PGM, and ALEX cannot outperform the B+-tree on disk even on the *COVID* dataset, which is considered to be an easy dataset as per [34]. The improvements from each search step attained in the main memory setting – e.g., a smaller search range in FITing-tree and PGM, or model-based search in ALEX – do not provide any tangible benefits for on-disk operations if they cannot reduce the number of fetched blocks.

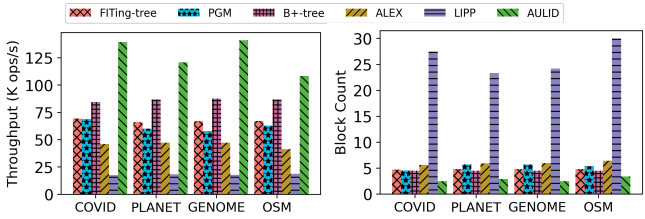

Figure 6: Throughput on Scan-Only Workload (W2).

The performance of the FITing-tree, PGM, and B+-tree is similar across all datasets. ALEX however has the worst performance on *OSM*. The performance of AULID and LIPP vary across different datasets; specifically, the performance is related to the *conflict degree* of a dataset, where a higher number of conflicts usually leads to a greater tree height, and in turn more fetched blocks.

*5.2.2 Scan-Only Workload.* Figure 6 summarizes the throughput and the average number of fetched blocks for the Scan-Only workload. In terms of throughput, AULID outperforms FITing-tree, PGM, B+-tree, ALEX, and LIPP by up to 2.11x, 2.44x, 1.65x, 3.04x, 7.94x, respectively. Just as in the Lookup-Only workload, the performance of the scans is determined by the number of fetched blocks.

To support a scan query, all indexes first initiate the search process for a lookup query to locate the start key in the search range, and then scan forward until reaching the end key. Consequently, better performance in Lookup-Only workloads yields better performance in Scan-Only workloads. Using the packed array in leaf nodes and links between siblings, the B+-tree and AULID reap the benefits from efficient lookup queries, and are the two top performing algorithms. In contrast, LIPP does not gain any benefits from lookup queries. LIPP only has one node type, where key-payload pairs, pointers to child nodes, and empty slots are all interleaved. Thus, when fetching the *next* item, LIPP may have to traverse multiple nodes. Since LIPP has a large fanout, there is a greater chance that these nodes are in different blocks. Also, we observe that the performance of ALEX decreases more quickly than the FITing-tree and PGM. This is because with a gapped array in the leaf node, ALEX uses a bitmap to indicate whether a slot is occupied, and thus incurs additional I/O cost when fetching it.

*5.2.3 Write-Only and Mixed Workloads.* Figure 7 shows the throughput for the workloads that include write operations. AULID is still the best performer across all workloads and datasets. The superiority of AULID is attributed to three reasons: (1) a lower latency to locate where a new key-payload should be inserted, i.e., benefiting from the best lookup performance; (2) a lower SMO overhead on leaf nodes with the B+-tree styled leaf node design; and (3) a lower SMO overhead for the inner nodes, and fewer SMOs required. Based on our design of the packed array and the two-layer B+-tree nodes, most new key-block pairs can be stored without creating new mixed nodes. In the tested datasets, no dataset required AULID creation of new mixed nodes. Other learned indexes, however, require more SMOs, e.g., on the Write-Only workload, ALEX and LIPP require 45,897 and 4.5M SMOs on *GENOME*, respectively.

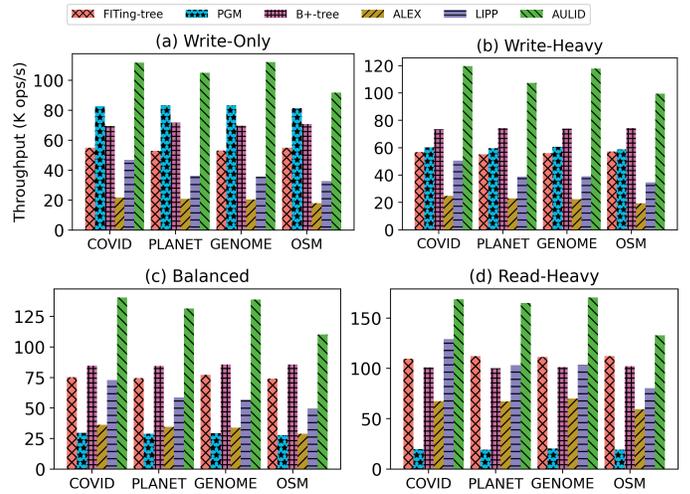

Figure 7: Throughput of Mixed Workloads (W3-W6).

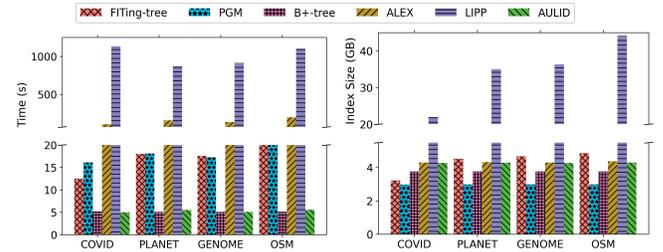

Figure 8: Comparison of Bulkload Time and Storage Usage.

From the other learned indexes, PGM outperforms other approaches on the Write-Only workload, but it performs worse when the ratio of read queries increases. Better insertion support stemming from the LSM tree [23] allows PGM to be competitive for write operations. However, since multiple files are maintained as static PGM indexes, PGM may access more than one file for a lookup query, which increases the I/O cost. The performance gain from a faster lookup time can benefit the workloads containing more reads, e.g., the FITing-tree and LIPP on the Read-Heavy workload. However, as the number of writes increases, the SMO overhead and the cost of updating statistics (for ALEX and LIPP) [34] can outweigh the benefits gained from faster lookups.

*5.2.4 Bulkloads.* Figure 8 reports the bulkloading time and the on-disk index size after bulkloading. When calculating index sizes, instead of only reporting the inner node sizes, we report the total size of the index file on disk. This ensures that the entire on-disk size is reported for a fair comparison in practice. In terms of bulkloading time, AULID is similar to the B+-tree, and both of them are significantly smaller than the other indexes. AULID also achieves similar storage cost to the B+-tree. The FITing-tree and LIPP have different storage sizes across different datasets. In the case of FITing-tree, harder datasets will create more leaf nodes (segments), and allocate additional buffers on disk for later key-payload pair insertions. For LIPP, a dataset with a larger degree of conflict will result in more nodes being created on disk, and in turn occupy more space. ALEX and LIPP have larger bulkloading times than the other methods due to model training, and more on-disk writes.

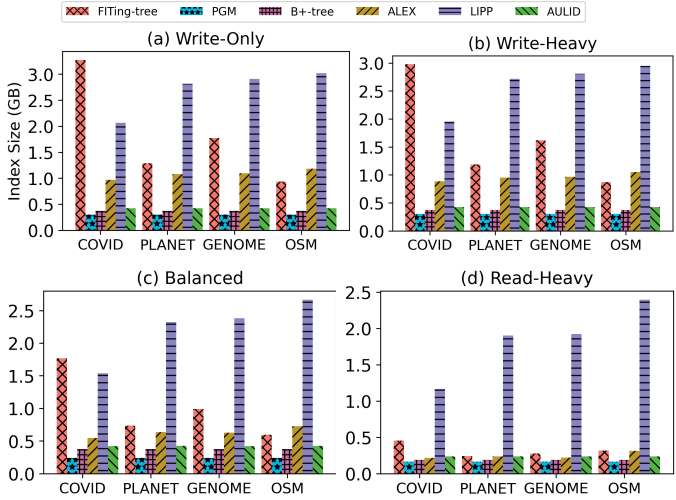

Figure 9: Storage Occupancy of Mixed Workloads (W3-W6).

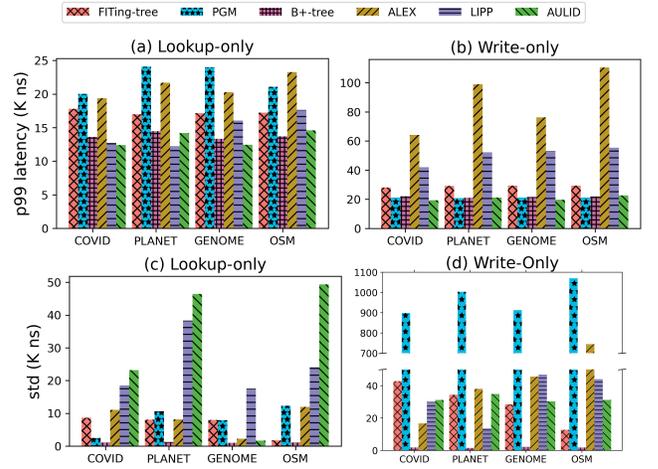

Figure 10: Tail Latency on Lookup-Only (W1) and Write-Only (W3) Workloads.

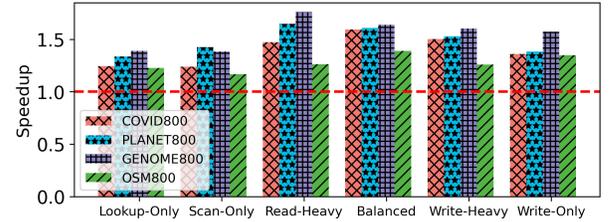

Figure 11: Throughput Speedup on Large Datasets.

*5.2.5 Index Size.* Figure 9 presents the storage occupancy of all indexes after finishing workloads comprising writes (**W3-W6**). Overall, AULID achieves similar storage overheads to the B+-tree across all workload and dataset combinations. Among the rest of the competitors, PGM has the smallest storage size. This is attributed to the LSM tree used in PGM to support arbitrary insertions, i.e. after an index has been merged, we can delete it from disk. For LIPP, a dataset with a higher degree of conflict usually has a larger storage cost due to the creation of additional nodes. The FITing-tree has a large space occupancy, regardless of the dataset or workload. For a hard dataset, re-segmenting a leaf node can generate many more leaf nodes compared to an easy one. This results in more buffers being created. For an easy dataset, a leaf node holds more items. Thus, each SMO operation writes more blocks on disk.

*5.2.6 Tail Latency.* To study the robustness of each index, in Figure 10 we report the p99 latency and standard deviation on the Lookup-Only and Write-Only workloads. Overall, AULID has the smallest p99 latency in the Lookup-Only workload. AULID, PGM, and B+-tree have similar p99 latencies for the Write-Only workload – all of which are better than the FITing-tree, ALEX, and LIPP. However, all learned indexes have a larger standard deviation than the B+-tree across both workloads. Due to an unbalanced tree structure of LIPP and ALEX, accessing some regions may issue more I/O requests for the Lookup-Only workload. When indexing only the largest key of each leaf node in the inner nodes, in a lookup, AULID may access more blocks to fetch the *next DATA* slot or read an extra block to locate the target leaf node as discussed in Section 4.2.3. PGM will periodically merge items into a larger index. Heavy SMOs for certain queries result in a larger latency in the Write-Only workload, which in turn results in a larger variance.

## 5.3 Scalability Test

In this section, we study the performance of AULID on large scale datasets using different workload types.

**Setting.** Since existing learned indexes perform worse than the B+-tree overall, in this section, we compare the scalability of AULID against the B+-tree only. To test the performance of AULID on datasets of different hardness, we include *OSM800* [21], and three other datasets of size 800M generated using the method from [34]. For each, we set the local hardness and global hardness to 4x of *COVID*, *PLANET*, and *GENOME* and name them as *COVID800*, *PLANET800*, and *GENOME800*, respectively.

For the Lookup-Only and Scan-Only workloads, we issue 800,000 queries over the index built on the entire dataset. Search keys are randomly sampled from the entire dataset. For the workloads that contain writes, we build an initial index containing 150M keys sampled from a dataset and issue a total of 50M operations, where the write ratio is the same as used for **W3-W6** in Section 5.1.3.

*5.3.1 Performance Speedup.* Figure 11 presents the throughput speedup of AULID compared to the B+-tree, across the four large datasets. AULID beats the B+-tree with up to 1.75x throughput gains on all tested workloads and datasets. AULID and B+-tree have the same leaf node layout. Due to a carefully designed inner node structure and an SMO mechanism to bound the tree height, AULID is more efficient when locating the target leaf node, and also benefits scans and writes.

The superiority of AULID on large datasets also comes from the smaller SMO overhead for write operations. When indexing the largest key for each leaf node of the learned model, AULID reduces the number of SMOs needed to reap the benefits of model-based search. Moreover, a packed array and a two-layer B+-tree hold the new key-block pairs without increasing the tree height (See Figure 4), while incurring only small update overheads.

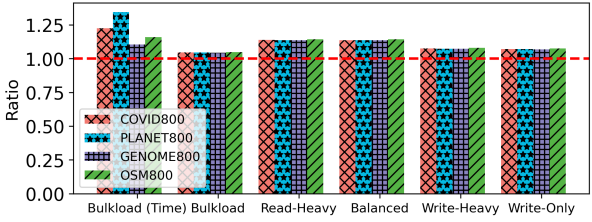

Figure 12: Bulkload Time and Storage Usage Compared to B+-tree on Large Datasets.

5.3.2 *Bulkload Time & Storage Usage.* We report the bulkloading time and index size for AULID and B+-tree in Figure 12. To build an index for 800M key-payload pairs on disk, the B+-tree takes around 20*s* and AULID is competitive at 27*s*. Both are much more efficient than other learned indexes, even on small datasets.

## 5.4 An In-depth Study of AULID Design

In this section, we study how the design of inner nodes meets our proposed design principles and address the two challenges aforementioned in Section 1.1. Typically, basic operations include the lookup and insertion, which in turn define the studied workload types (**W1, W3-W6**) and are also the key step in **W2**. Specifically, we first study the impact of the AULID design on these two operations, and then investigate the impact of the adjustment strategy proposed in Section 4.4.

5.4.1 *Impact of Different Designs on Lookup-Only Workloads.* To study the effectiveness of the proposed design, we compare AULID against LIPP-B+ – an approach which directly adopts LIPP as the inner nodes, and organizes the leaf nodes in the same vein as B+-tree. We report the throughput of the Lookup-only Workload (**W1**) in Table 2. Across all datasets, AULID outperforms LIPP-B+ and fetches fewer blocks.

Table 2: Throughput Comparison of AULID and LIPP-B+ on Lookup-Only Workload (W1).

| Metric | Index | COVID | PLANET | GENOME | OSM |
|---|---|---|---|---|---|
| Throughput | LIPP-B+ | 158,489 | 133,404 | 153,851 | 104,659 |
|  | AULID | 164,897 | 141,543 | 163,825 | 123,749 |
| Blocks | LIPP-B+ | 2.15 | 2.72 | 2.25 | 3.30 |
|  | AULID | 2.07 | 2.50 | 2.07 | 2.96 |

The performance of AULID is attributed to our *read optimization strategies* (Section 4.2.3), *packed array design*, and *two-layered B+-tree nodes* (Section 3.3.2). The first helps reduce the number of blocks being fetched, and the latter two help reduce the tree height.
**Benefits of Read Optimizations.** As discussed in Section 4.2.3, in AULID, there are two cases that may incur additional I/O costs: *Case 1* - located in the predecessor of the target leaf node and require an additional block being fetched; *Case 2* - located in a *NULL* slot but need to scan forward until the next *DATA* slot is found. For the first case, AULID scans forward (**ScanFward**) to determine whether the current block has another *DATA* slot. For the second case, AULID

Table 3: Extra Fetched Blocks under Different Optimizations

| Dataset | w/o Opt. | Fulfill | ScanFward | Fulfill & ScanFward |
|---|---|---|---|---|
| COVID | 26,107 | 18,337 | 9,266 | 277 |
| PLANET | 59,711 | 52,619 | 30,090 | 21,027 |
| GENOME | 47,229 | 40,727 | 9,456 | 710 |
| OSM | 30,148 | 22,368 | 23,232 | 14,924 |

fulfills (**Fulfill**) the empty slot with the next *DATA* slot during a bulkloading process.

With the **Fulfill** optimization, AULID avoids accessing extra blocks in *Case 2*. Thus, all of the fetched extra blocks in the **Fullfill** optimization are stemming from *Case 1* (the third column in Table 3). From Table 3, we observe that most additional fetched blocks are from *Case 2*.

With the **ScanFward** optimization, AULID significantly reduces the number of extra fetched blocks for *COVID*, *PLANET*, and *GENOME*. With the **Fulfill** optimization, AULID avoids fetching extra blocks in *Case 1*.

When enabling both operations, AULID can reduce the number of extra fetched blocks by at least 50%, particularly for *COVID* and *GENOME*. By default, we only enable the **ScanFward** optimization, which can reduce 0.08, 0.15, 0.18, and 0.03 blocks per query for *COVID*, *PLANET*, *GENOME*, and *OSM*, respectively. Thus, improvements in AULID on *COVID* and *GENOME* in Table 2 are produced by **ScanFward** optimization only.

Table 4: Impact of Packed Array and Two-Layer B+-tree on the Average *DATA* Slot Height and Storage.

| Metric | Index | COVID | PLANET | GENOME | OSM | OSM800 |
|---|---|---|---|---|---|---|
| Avg. Height | LIPP-B+ | 1.00 | 1.60 | 1.01 | 2.29 | 2.28 |
|  | AULID | 1.00 | **1.36** | **1.00** | **1.83** | **1.93** |
| Storage (GB) | LIPP-B+ | 4.27 | 4.47 | 4.28 | 5.11 | 19.51 |
|  | AULID | 4.27 | **4.28** | **4.27** | **4.29** | **15.77** |

**Benefits of Data Structure Design.** The design of the packed array and the two-layer B+-tree node in AULID helps further reduce the number of fetched blocks for the *PLANET* and *OSM* datasets due to the lower tree height (as compared to LIPP-B+).

Table 4 reports the average node heights after a bulkload on *COVID* and *GENOME*. AULID and LIPP-B+ both have the minimal average node height, where most *DATA* slots are located in the first level. However, for datasets with a larger *conflict degree*, AULID has a smaller average node height. This is because LIPP-B+ creates more nodes to eliminate the number of conflicts in each dataset with a larger *conflict degree*, which leads to a greater height. A lower tree height can be achieved with the packed data structures discussed above, and thus AULID requires less storage space than LIPP-B+ on the hard datasets as shown in Table 4.

Table 5: Throughput Comparison of AULID and LIPP-B+ on Write-Only Workload (W3)

| Metric | Index | COVID | PLANET | GENOME | OSM |
|---|---|---|---|---|---|
| Throughput | LIPP-B+ | 111,017 | 100,430 | 109,631 | 76,865.4 |
|  | AULID | 111,669 | 104,816 | 107,661 | 91,707.1 |

5.4.2 *Packed Array, Two-Layer B+-tree Nodes, and B+-tree Styled Leaf Nodes for Write-Only Workloads.* Table 5 reports the throughput for the Write-Only workload (**W3**). We observe that AULID and LIPP-B+ have similar performance for *COVID*, *PLANET*, and *GENOME*, but AULID outperforms LIPP-B+ by 1.19x on *OSM*.

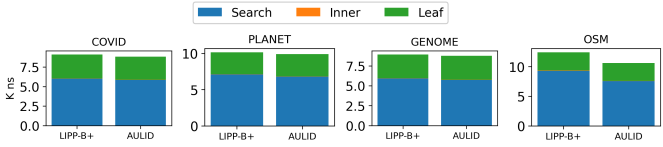

Figure 13: Latency Breakdown of Write-Only Workload (W3).

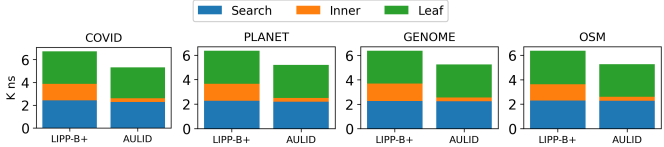

Figure 14: Latency Breakdown of Append-Only Workload.

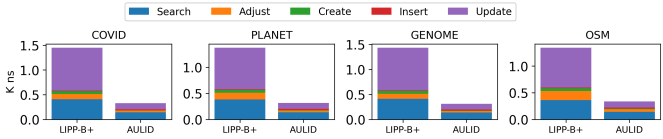

Figure 15: Inner Nodes' Latency Breakdown of Append-Only Workload.

To further understand the performance, we break down the insertion process into three steps: (1) the search step (**Search**) to find the leaf node that will hold the new key-payload pair, (2) the insertion step on a leaf node (**Leaf**), and (3) the update step in the inner nodes (**Inner**). From Figure 13, we can see that the main overhead is brought by the first two steps. The overhead from indexing the new key-block pairs in the inner nodes (**Inner**) is negligible. AULID and LIPP-B+ update the inner nodes when a leaf node is split. When compared against LIPP, both have fewer SMO operations. For example, on *GENOME*, AULID and LIPP-B+ only require 49,038 SMOs on the leaf nodes, where the leaf nodes must be split; in contrast, LIPP requires 4.6M SMOs and most of them are caused by creating LIPP nodes to eliminate conflicts. On *OSM*, AULID has a more efficient **Search** step, which contributes to the higher throughput, as compared to LIPP-B+. The design of the packed array and the two-layer B+-tree nodes in AULID consistently results in shorter paths to leaf nodes, and hence smaller search cost.

Table 6: Throughput Comparison of AULID and LIPP-B+ on Append-Only Workload

| Metric | Index | COVID | PLANET | GENOME | OSM |
|---|---|---|---|---|---|
| Throughput | LIPP-B+ | 144,432 | 153,060 | 152,633 | 153,214 |
| | AULID | 182,732 | 188,015 | 187,886 | 184,051 |

**Hot Region Insertions.** Another potential problem with LIPP are insertions in a hot region which occur in the inner nodes. This produces a high number of conflicts and triggers additional SMOs. We use here an Append-Only workload to analyze this case. The throughput and latency breakdown for this case are shown in Table 6 and Figure 14, respectively. To support new key-payload pairs, AULID and LIPP-B+ index the last leaf node in the meta-node. Thus, they have similar performance on the **Search** and **Leaf** in Figure 6. AULID has a lower latency when updating the inner nodes (**Inner**) across all datasets.

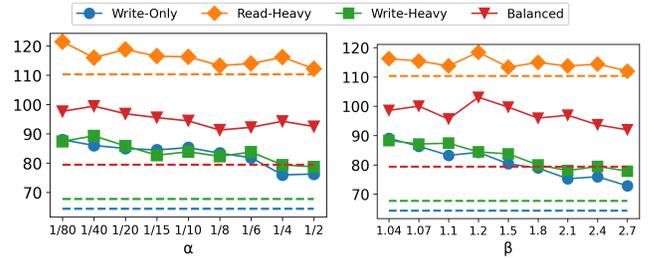

Figure 16: Throughput under Different Settings. The dashed line indicates the throughput of the corresponding workload with the same color without any adjustments.

To study the overhead of the inner nodes, we further break down the process of the inner nodes into five steps: (1) search to find a slot to hold the new key-block pair (**Search**), (2) create a new node or convert the node type (**Create**), (3) insert the key-block pair into a *DATA* slot (**Insert**), (4) adjust the tree structure (**Adjust**), and (5) update the statistics for later adjustment operations (**Update**).

In Figure 15, we observe that, except for the **Insert**, AULID has a lower latency than LIPP-B+. The packed array and two-layer B+-tree node design of AULID yield a shorter path to target slots holding the new key-block pairs, which contributes to lower search and update times. Moreover, conflicts in the Append-Only workload require more operations when creating LIPP nodes and adjusting tree structures in LIPP-B+. For example, AULID only requires 1,163 and 37 SMOs in creating new nodes and adjusting the tree structure, while LIPP-B+ requires 38,837 and 1,625 SMOs, respectively.

5.4.3 *Adjustment Study.* Last, we study the effectiveness of index adjustments and provide an analysis on the parameters in AULID – $\alpha$ and $\beta$ presented in Algorithm 2. We use the *OSM* dataset, and build an initial index using 50M keys. Then, another 50M queries are issued with different write ratios, just as in **W3-W6**. To study the impact of $\alpha$ and $\beta$, we set $\beta = 1.07$ and $\alpha = 0.0025$ as the default. Using small default values makes the corresponding condition true (Line 2 in Algorithm 2), and hence we can study the impact of other parameters in isolation. Figure 16 illustrates the throughput when using different settings. We also report the throughput without adjusting the index (the dashed lines). In Figure 16, we observe: (1) The index adjustment in AULID significantly improves the throughput, especially for workloads with write operations. (2) Larger values of $\alpha$ and $\beta$ usually result in worse performance, and the workloads with more writes are more sensitive to these two parameters. If index is not adjusted, certain regions in the inner nodes can result in longer paths to the leaf nodes. This incurs more reads to fetch leaf nodes and more writes to update statistics on disk. We find that the default values of $\alpha$ (0.05) and $\beta$ (1.2) in AULID result in good performance across all tested workloads.

## 6 CONCLUSION

In this paper, we propose AULID, a novel simple yet efficient on-disk learned index. In contrast to in-memory indexes, I/O cost is the main bottleneck for disk-resident indexes. Toward that end, we propose five principles to build a learned index on disk, focusing on reducing the I/O cost. We carefully craft the index structure, propose the query processing algorithms, and introduce an index adjustment mechanism to meet the proposed principles. AULID outperforms all of the baselines across a wide range of settings.

Our evaluation shows that AULID has competitive storage cost to the B+-tree (the smallest of the alternatives), and achieves up to 2.11x, 8.63x, 1.72x, 5.51x, and 8.02x higher throughput than the FITing-tree, PGM, B+-tree, ALEX, and LIPP respectively.

# REFERENCES

[1] Hussam Abu-Libdeh, Deniz Altinbüken, Alex Beutel, Ed H. Chi, Lyric Doshi, Tim Kraska, Xiaozhou Li, Andy Ly, and Christopher Olston. 2020. Learned Indexes for a Google-scale Disk-based Database. *CoRR* abs/2012.12501 (2020).

[2] Andreas Behrend, Anton Dignös, Johann Gamper, Philip Schmiegelt, Hannes Voigt, Matthias Rottmann, and Karsten Kahl. 2019. Period Index: A Learned 2D Hash Index for Range and Duration Queries. In *SSTD*. ACM, 100–109.

[3] Yifan Dai, Yien Xu, Aishwarya Ganesan, Ramnatthan Alagappan, Brian Kroth, Andrea C. Arpaci-Dusseau, and Remzi H. Arpaci-Dusseau. 2020. From WiscKey to Bourbon: A Learned Index for Log-Structured Merge Trees. In *OSDI*. 155–171.

[4] Jialin Ding, Umar Farooq Minhas, Jia Yu, Chi Wang, Jaeyoung Do, Yinan Li, Hantian Zhang, Badrish Chandramouli, Johannes Gehrke, Donald Kossmann, David B. Lomet, and Tim Kraska. 2020. ALEX: An Updatable Adaptive Learned Index. In *SIGMO*. 969–984.

[5] Jialin Ding, Vikram Nathan, Mohammad Alizadeh, and Tim Kraska. 2020. Tsunami: A Learned Multi-dimensional Index for Correlated Data and Skewed Workloads. *Proc. VLDB Endow.* 14, 2 (2020), 74–86.

[6] Paolo Ferragina, Fabrizio Lillo, and Giorgio Vinciguerra. 2020. Why Are Learned Indexes So Effective?. In *ICML (Proceedings of Machine Learning Research, Vol. 119)*. PMLR, 3123–3132.

[7] Paolo Ferragina and Giorgio Vinciguerra. 2020. The PGM-index: a fully-dynamic compressed learned index with provable worst-case bounds. *Proc. VLDB Endow.* 13, 8 (2020), 1162–1175.

[8] Alex Galakatos, Michael Markovitch, Carsten Binnig, Rodrigo Fonseca, and Tim Kraska. 2019. FITing-Tree: A Data-aware Index Structure. In *SIGMOD*. 1189–1206.

[9] Laura M. Haas, Michael J. Carey, Miron Livny, and Amit Shukla. 1997. Seeking the Truth About ad hoc Join Costs. *VLDB J.* 6, 3 (1997), 241–256.

[10] Ali Hadian and Thomas Heinis. 2019. Interpolation-friendly B-trees: Bridging the Gap Between Algorithmic and Learned Indexes. In *EDBT*. OpenProceedings.org, 710–713.

[11] Ali Hadian and Thomas Heinis. 2021. Shift-Table: A Low-latency Learned Index for Range Queries using Model Correction. In *EDBT*. OpenProceedings.org, 253–264.

[12] Andreas Kipf, Dominik Horn, Pascal Pfeil, Ryan Marcus, and Tim Kraska. 2022. LSI: a learned secondary index structure. In *aiDM@SIGMOD*. ACM, 4:1–4:5.

[13] Andreas Kipf, Ryan Marcus, Alexander van Renen, Mihail Stoian, Alfons Kemper, Tim Kraska, and Thomas Neumann. 2020. RadixSpline: a single-pass learned index. In *aiDM@SIGMOD*. ACM, 5:1–5:5.

[14] Tim Kraska, Alex Beutel, Ed H. Chi, Jeffrey Dean, and Neoklis Polyzotis. 2018. The Case for Learned Index Structures. In *SIGMOD*. 489–504.

[15] Pengfei Li, Yu Hua, Jingnan Jia, and Pengfei Zuo. 2021. FINEdex: A Fine-grained Learned Index Scheme for Scalable and Concurrent Memory Systems. *Proc. VLDB Endow.* 15, 2 (2021), 321–334.

[16] Pengfei Li, Hua Lu, Qian Zheng, Long Yang, and Gang Pan. 2020. LISA: A Learned Index Structure for Spatial Data. In *SIMGOD*. 2119–2133.

[17] Anisa Llaveshi, Utku Sirin, Anastasia Ailamaki, and Robert West. 2019. Accelerating B+ tree search by using simple machine learning techniques. In *Proceedings of the 1st International Workshop on Applied AI for Database Systems and Applications*.

[18] Baotong Lu, Jialin Ding, Eric Lo, Umar Farooq Minhas, and Tianzheng Wang. 2021. APEX: A High-Performance Learned Index on Persistent Memory. *Proc. VLDB Endow.* 15, 3 (2021), 597–610.

[19] Chaohong Ma, Xiaohui Yu, Yifan Li, Xiaofeng Meng, and Aishan Maoliniyazi. 2022. FILM: a Fully Learned Index for Larger-than-Memory Databases. *Proc. VLDB Endow.* 16, 3 (2022), 561–573.

[20] Marcel Maltry and Jens Dittrich. 2022. A Critical Analysis of Recursive Model Indexes. *Proc. VLDB Endow.* 15, 5 (2022), 1079–1091.

[21] Ryan Marcus, Andreas Kipf, Alexander van Renen, Mihail Stoian, Sanchit Misra, Alfons Kemper, Thomas Neumann, and Tim Kraska. 2020. Benchmarking Learned Indexes. *Proc. VLDB Endow.* 14, 1 (2020), 1–13.

[22] Vikram Nathan, Jialin Ding, Mohammad Alizadeh, and Tim Kraska. 2020. Learning Multi-Dimensional Indices. In *SIGMOD*. ACM, 985–1000.

[23] Patrick E. O'Neil, Edward Cheng, Dieter Gawlick, and Elizabeth J. O'Neil. 1996. The Log-Structured Merge-Tree (LSM-Tree). *Acta Informatica* 33, 4 (1996), 351–385.

[24] Joseph O'Rourke. 1981. An On-Line Algorithm for Fitting Straight Lines Between Data Ranges. *Commun. ACM* 24, 9 (1981), 574–578.

[25] Sachith Gopalakrishna Pai, Michael Mathioudakis, and Yanhao Wang. 2022. Towards an Instance-Optimal Z-Index. In *4th International Workshop on Applied AI for Database Systems and Applications (AIDB@ VLDB2022)*.

[26] Varun Pandey, Alexander van Renen, Andreas Kipf, Jialin Ding, Ibrahim Sabek, and Alfons Kemper. 2020. The Case for Learned Spatial Indexes. In *AIDB@VLDB*.

[27] Jianzhong Qi, Guanli Liu, Christian S. Jensen, and Lars Kulik. 2020. Effectively Learning Spatial Indices. *Proc. VLDB Endow.* 13, 11 (2020), 2341–2354.

[28] Danilo Jimenez Rezende and Shakir Mohamed. 2015. Variational Inference with Normalizing Flows. In *ICML*, Vol. 37. JMLR.org, 1530–1538.

[29] Benjamin Spector, Andreas Kipf, Kapil Vaidya, Chi Wang, Umar Farooq Minhas, and Tim Kraska. 2021. Bounding the Last Mile: Efficient Learned String Indexing. *CoRR* abs/2111.14905 (2021).

[30] Mihail Stoian, Andreas Kipf, Ryan Marcus, and Tim Kraska. 2021. PLEX: Towards Practical Learned Indexing. *CoRR* abs/2108.05117 (2021).

[31] Chuzhe Tang, Youyun Wang, Zhiyuan Dong, Gansen Hu, Zhaoguo Wang, Minjie Wang, and Haibo Chen. 2020. XIndex: a scalable learned index for multicore data storage. In *PPoPP*. ACM, 308–320.

[32] Youyun Wang, Chuzhe Tang, Zhaoguo Wang, and Haibo Chen. 2020. SIndex: a scalable learned index for string keys. In *APSys '20: 11th ACM SIGOPS Asia-Pacific Workshop on Systems, Tsukuba, Japan, August 24-25, 2020*. ACM, 17–24.

[33] H.-S. Philip Wong, Simone Raoux, SangBum Kim, Jiale Liang, John P. Reifenberg, Bipin Rajendran, Mehdi Asheghi, and Kenneth E. Goodson. 2010. Phase Change Memory. *Proc. IEEE* 98, 12 (2010), 2201–2227.

[34] Chaichon Wongkham, Baotong Lu, Chris Liu, Zhicong Zhong, Eric Lo, and Tianzheng Wang. 2022. Are Updatable Learned Indexes Ready? *Proc. VLDB Endow.* 15, 11 (2022), 3004–3017.

[35] Jiacheng Wu, Yong Zhang, Shimin Chen, Yu Chen, Jin Wang, and Chunxiao Xing. 2021. Updatable Learned Index with Precise Positions. *Proc. VLDB Endow.* 14, 8 (2021), 1276–1288.

[36] Shangyu Wu, Yufei Cui, Jinghuan Yu, Xuan Sun, Tei-Wei Kuo, and Chun Jason Xue. 2022. NFL: Robust Learned Index via Distribution Transformation. *Proc. VLDB Endow.* 15, 10 (2022), 2188–2200.

[37] Yingjun Wu, Jia Yu, Yuanyuan Tian, Richard Sidle, and Ronald Barber. 2019. Designing Succinct Secondary Indexing Mechanism by Exploiting Column Correlations. In *SIGMOD*. ACM, 1223–1240.

[38] Geoffrey X. Yu, Markos Markakis, Andreas Kipf, Per-Åke Larson, Umar Farooq Minhas, and Tim Kraska. 2022. TreeLine: An Update-in-Place Key-Value Store for Modern Storage. *Proc. VLDB Endow.* 16, 1 (2022), 99–112.

[39] Jiaoyi Zhang and Yihan Gao. 2022. CARMI: A Cache-Aware Learned Index with a Cost-based Construction Algorithm. *Proc. VLDB Endow.* 15, 11 (2022), 2679–2691.

[40] Xuanhe Zhou, Luyang Liu, Wenbo Li, Lianyuan Jin, Shifu Li, Tianqing Wang, and Jianhua Feng. 2022. AutoIndex: An Incremental Index Management System for Dynamic Workloads. In *ICDE*. IEEE, 2196–2208.